\documentclass{emulateapj}

\usepackage{natbib}
\usepackage{graphicx}
\usepackage{psfig}

\newcommand{\tnm}{\tablenotemark}
\newcommand{\tnt}{\tablenotetext}

\newcommand{\ergs}{ergs s$^{-1}$}
\newcommand{\flux}{ergs cm$^{-2}$ s$^{-1}$}

\newcommand{\xmm}{{\it XMM}}

\newcommand{\chandra}{{\it Chandra}}
\newcommand{\asca}{{\it ASCA}}
\newcommand{\rxte}{{\it RXTE}}
\newcommand{\rosat}{{\it ROSAT}}
\newcommand{\bepposax}{{\it BeppoSAX}}
\newcommand{\ginga}{{\it Ginga}}

\newcommand{\rxj}{RX J0059.2--7138}
\newcommand{\xtej}{XTE J0111.2--7317}
\newcommand{\fouru}{4U 1626--67}

\newcommand{\tout}{\theta_{\rm out}}
\newcommand{\tin}{\theta_{\rm in}}
\newcommand{\rout}{r_{\rm out}}
\newcommand{\rin}{r_{\rm in}}

\newcommand{\im}{\item}

\newcommand{\dgr}{$^{\circ}$}

\newcommand{\W}{\hphantom{0}}

\begin{document}

\slugcomment{Accepted for publication in the Astrophysical Journal}

\title{Pulse-phase spectroscopy of SMC X-1 with {\it Chandra} \\
and {\it XMM-Newton}:
reprocessing by a precessing disk?}
\shorttitle{SPECTROSCOPY OF SMC X-1}
\shortauthors{HICKOX \& VRTILEK}
\author{R.C. Hickox}
\author{S.D. Vrtilek}
\affil{Harvard-Smithsonian Center for Astrophysics, 60 Garden Street,
 Cambridge, MA 02138}

\begin{abstract}
We present pulse-phase X-ray spectroscopy of the high-mass X-ray
pulsar SMC X-1 from five different epochs, using {\it Chandra X-ray
Observatory} ACIS-S and {\it XMM-Newton} EPIC-pn data.  The X-ray
spectrum consistently shows two distinct components, a hard power law
and a soft blackbody with $kT_{\rm BB}\sim 0.18$ keV.  For all five
epochs the hard component shows a simple double-peaked pulse shape,
and also a variation in the power law slope, which becomes harder at
maximum flux and softer at minimum flux.  For the soft component, the
pulse profile changes between epochs, in both shape and phase relative
to the power law pulses.  The soft component is likely produced by
reprocessing of the hard X-ray pulsar beam by the inner accretion
disk.  We use a model of a twisted inner disk, illuminated by
the rotating X-ray pulsar beam, to simulate pulsations in the soft
component due to this reprocessing.  We find that for some disk and
beam geometries, precession of an illuminated accretion disk can
roughly reproduce the observed long-term changes in the
soft pulse profiles.

\end{abstract}

\section{Introduction}
The general picture of accretion onto X-ray pulsars, consisting of a
flow in a disk to the magnetosphere and along the dipole field to the
magnetic poles of the neutron star, has been known for decades
\citep[see][and references therein]{naga89}.  However, despite
extensive study, some important aspects of this process remain
mysterious.  In particular, the details of accretion near the
magnetosphere, where the neutron star's magnetic field begins to
dominate the flow, is a complex and challenging problem \citep{elsn77,ghos77,ghos78}.

\subsection{Superorbital variation in X-ray pulsars}

One way to explore an accretion flow observationally is with systems where
the shadow of the disk causes the X-ray source to vary with
time.  There are now $\sim$20 known accreting binary systems that
exhibit superorbital periods, some of which are caused by shadowing of
the source flux by an accretion disk \citep[see][and references therein]{clar03}.  Of these systems, a small
group of X-ray pulsars (SMC X-1, LMC X-4 and Hercules X-1) show luminous,
pulsing X-ray emission which allows for study of the accretion flow in
unusual detail.  These systems each show
a 30--60 d superorbital variation, attributed to a warped, precessing
accretion disk \citep{tana72, jone76, katz73, lang81,levi96,grub84,wojd98}.

In the low-mass system Her X-1, the precessing disk has been
studied extensively.  Regular 35-day variations have been observed in
the X-ray flux, pulse profiles, and spectrum, as well as optical
and UV features \citep[e.g.,][and references therein]{vrti94,deet98}.  In
particular, occulation by the disk produces variations in the
pulse profile shapes, which can be used to infer the overall warped
disk geometry \citep{leah02,scot00}.

SMC X-1 and LMC X-4 differ from Her X-1 in that they have high-mass (O
and B star) companions, hence we expect the accretion mechanism to be
somewhat different due to the presence of powerful stellar winds.
However, the high X-ray luminosities of SMC X-1 and LMC X-4
($>10^{38}$ \ergs) require Roche lobe overflow in addition to
accretion from the wind, and there is evidence for persistent,
warped accretion disks. 

For example, in LMC X-4, precession of the disk is believed to cause changes in
the X-ray power-law slope and flourescent iron line flux as a function
of superorbital phase \citep{naik03, naik02}, as well as variations in
the ultraviolet spectra \citep{prec02}.  For SMC X-1, observations of
X-ray spectra \citep{wojd98} and photoionized lines in the UV and
X-ray \citep{vrti01, vrti05} give evidence for shadowing of the
central neutron star by the precessing disk.  From the observed
line features, \citet{vrti05} conclude that the occulation of the source is caused by the inner
regions of the disk, near the magnetospheric
radius $\sim$$10^8$ cm.

\subsection{Disk reprocessing and the soft X-ray emission}

In Her X-1, reprocessing of the hard X-rays from the neutron star by
the inner accretion disk produces a soft ($< 1$ keV) X-ray component that pulses
out of phase with the hard ($> 2$ keV) pulses \citep[e.g.][]{endo00,
rams02}.  Thus the X-ray beam from the neutron star acts as a
``flashlight'' that illuminates the inner disk (Fig. 1).

\begin{figure}
\plotone{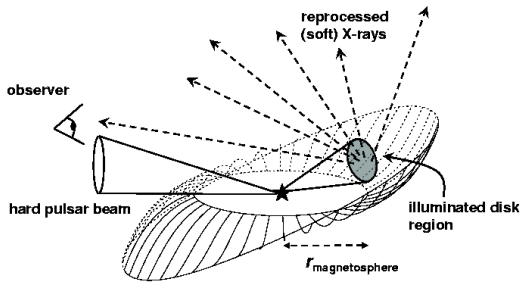}
\caption{Schematic of disk reprocessing in X-ray pulsars.  The hard (power law
spectrum) X-ray beam
sweeps around and illuminates the accretion disk, which re-radiates a
soft X-ray (blackbody) component.  The observer sees pulses from both
the power law and blackbody components, but these differ in pulse shape and phase.}
\end{figure}

Many luminous X-ray pulsars, including LMC X-4 and SMC X-1, have
similar soft spectral components that are likely also caused by disk
reprocessing \citep{hick04}.  We expect the flux of such a soft
component to pulsate as the pulsar beam sweeps around and
illuminates the disk.  The soft pulses will have the same period as
those from the hard X-ray beam, but will have a different shape and
relative phase.  Moreover, precession of the accretion disk will
change the observer's view of the inner region, and so the shape and
phase of the reprocessed component should vary with superorbital
period.  Hints of such an effect have been seen in Her
X-1: \citet{zane04} showed using \xmm\ data that the phase offset between
the soft (0.3--0.7 keV) and hard (3--10 keV) pulse profiles vary with
time, and appears to be correlated with precession of the disk.  A
soft component that pulsates separately from the hard component has
also been seen in LMC X-4 \citep{naik04} and SMC X-1 \citep{paul02,
naik04a}.

\subsection{SMC X-1}

SMC X-1 was first identified as a point source in {\it Uhuru}
observations by \citet{leon71}.  Eclipses were discovered by
\citet{schr72} and X-ray pulsations by \citet{luck76}.  The optical
counterpart is Sk 160, a B0 I supergiant \citep{webs72, lill73}. The
orbit is close to circular, with orbital period $\sim$3.9 days and
pulse period 0.7 s. The source has a high X-ray luminosity, $L_{\rm X}
\sim 3 \times 10^{38}$ \ergs, in the bright state.

In SMC X-1 the period of the superorbital cycle oscillates from 40 to
60 days \citep{wojd98}, a variation that may itself be periodic on the
order of years \citep{clar03}. The high-state, out-of-eclipse X-ray
spectrum is usually fitted as a cutoff power law with
$\Gamma \sim 0.9$, an iron line at $\sim$6.4 keV, and a soft component
modeled as a broken power law, thermal bremsstrahlung, or a blackbody
with $kT_{\rm BB} \sim 0.16$ keV. \citet{paul02} fit four different
models to the soft component and found that all gave satisfactory
fits.  They found that the soft component flux pulses roughly
sinusoidally, out of phase with the sharper-peaked pulses from the
power law.  \citet{neil04} created hard and soft pulse profiles for \rosat,
\chandra, and \xmm\ data of SMC X-1, and found that for different
epochs the the soft pulse shapes vary while the hard pulses remain similar.

In this paper we seek to explore the variation in these spectral
components further, by performing pulse-phase spectroscopy for SMC X-1
at several epochs.  We have also constructed a model to simulate hard
and soft pulse profiles that would be produced by an X-ray pulsar beam
illuminating a warped, precessing accretion disk, and compare the
results with observations.

\section{Observations and data reduction}

\begin{figure}
\plotone{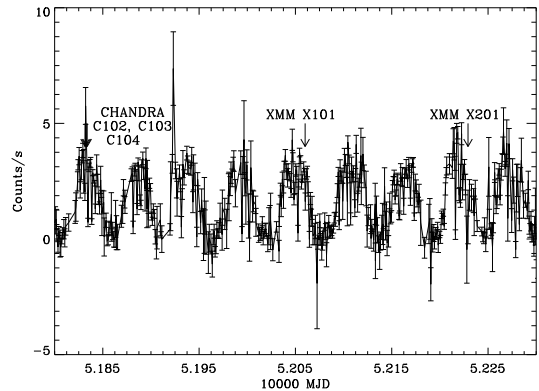}
\caption{\textit{RXTE} ASM lightcurve for SMC X-1 with arrows designating
several observations. Quick-look results as provided by the \textit{RXTE}/ASM team.}
\end{figure}

\begin{figure}
\plotone{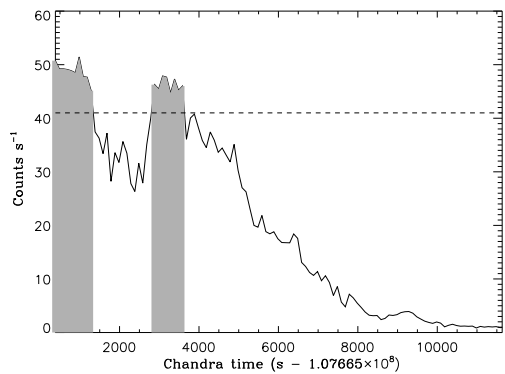}
\caption{XMM EPIC-pn light curve (0.15--12 keV) for observation X101,
showing entrance to eclipse.  Times with flux $>41$ cts s$^{-1}$
(shaded) were
included in the spectral analysis.}
\end{figure}

We have used three \chandra\ and two \xmm\ observations of SMC X-1,
taken in the high state of the superorbital cycle.  Details of the
observations are given in Table 1. Pulse profile analysis for these
data are presented in \citet{neil04}, and we will use their
abbreviations to denote the observations (C102, X101, etc.). Fig. 2
shows the observations on the long-term light curve of SMC X-1 from
the All Sky Monitor (ASM) on \rxte.  From this we estimate the
superorbital phases given in Table 1, taking $\phi_{\rm SO}=0$ at
the start of the superorbital high state.  These estimates have
uncertainty of $\pm$ a few days or $\phi_{\rm SO} \sim 0.05$, because
the superorbital intensity variation in SMC X-1 is quasi-periodic in
nature and hence does not have a well-defined ephemeris.

\begin{deluxetable*}{lcccccccc}
\tabletypesize{\scriptsize}
\tablecaption{\chandra\ and \xmm\ observations of SMC X-1}
\tablewidth{0pt}
\tablehead{
\colhead{Observation ID}  &
\colhead{Mission}  &
\colhead{Start Time}  & 
\colhead{Total}  & 
\colhead{$\phi_{\rm orb}$\tnm{b}} &
\colhead{$\phi_{\rm SO}$} &
\multicolumn{3}{c}{Fluxes ($10^{-10}$ \flux)\tnm{c}} \\
\colhead{(Reference)}  &
\colhead{}  &
\colhead{(JD-2400000)}  &
\colhead{Exp. (ks)\tnm{a}} & 
\colhead{} &
\colhead{} &
\colhead{Power law} &
\colhead{Blackbody} &
\colhead{Total}}
\startdata
400102 (C102)      &  \textit{Chandra}  &  51832.17  &  \W6.2 & 0.19 
& 0.16 & 9.3 & 0.9 & 10.2\\
400103 (C103)      &  \textit{Chandra}  &  51833.34  &  \W6.1 & 0.49 
& 0.16 & 9.1 & 1.6 & 10.7 \\
400104 (C104)      &  \textit{Chandra}  &  51834.31  &  \W6.5 & 0.74 
& 0.16 & 9.2 & 1.4 & 10.6 \\
0011450101 (X101)  &  \textit{XMM}      &  52060.59  &  27.9 & 0.89
& 0.37 & 6.5 & 0.4 & 6.9   \\
0011450201 (X201)  &  \textit{XMM}      &  52229.65  &  \W1.6 & 0.34 
& 0.42 & 10.1 & 1.3 & 11.4
\enddata
\tnt{a}{Clean, out-of-eclipse exposure.}
\tnt{b}{Orbital phase at start of exposure.}
\tnt{c}{Unabsorbed fluxes in the range 0.5--10 keV.}
\end{deluxetable*}

The \chandra\ observations were each $\sim$6 ks in duration, and they
were taken with the ACIS-S3 chip in Continuous Clocking (CC) mode, in
which charge is continually read out down the chip.  This allows for
3.5 ms time resolution, but one dimension of spatial information is
lost so that the X-ray image is simply a straight line. There are no
other bright sources within the ACIS field of view near SMC X-1, so we
can determine source and background regions from the 1-D image and
extract events normally.  For each observation the source region was a
rectangle of size $0\farcs 49 \times 1\farcs 97$ centered on the
source.  Background events came from two rectangles $0\farcs 49 \times
9\farcs 84$ on either side of the source.

For the \xmm\ observations we have used data from the EPIC-pn CCD
camera, which has the highest time resolution of the EPIC cameras.
X201 had a clean exposure of $\sim$27 ks in duration, and was taken in
Small Window mode, which has 6 ms time resolution.  For this
observation we have extracted source events from a circle of radius
$51\farcs 2$.  X101 was taken in Full Window mode, which has 73.5 ms
time resolution.  While the total exposure for X101 was 40 ks, the
bulk of this observation was during full eclipse, with only the first
few ks showing substantial flux.  \citet{neil04} analyzed the first
4.9 ks of data, which consisted of the entire entrance into eclipse.
However, they found that a partial covering absorption model was
needed to explain the X-ray spectrum, perhaps due to occultation by the
outer layers of the star.  To avoid this complication, we restricted
the present analysis to the 1.6 ks of data for which the 0.15-12 keV
count rate was $>41$ counts s$^{-1}$ (see Fig. 3).  The background
surface brightness for the \xmm\ data was $<0.1$\% of the source
brightness, so we did not perform background subtraction.

\section{Data Analysis}

\subsection{Dealing with event pileup}

In high count-rate data such as these, the X-rays measured by CCD detectors are
affected by \textit{pileup}, caused when two or more photons are
coincident on one event-detection cell during one
time-resolution element, or frame time \citep{davi01}.  In this case
the detector cannot distinguish the multiple events and instead
detects a single event with a pulse height roughly equal to the sum of
the pulse heights from the individual photons.  Pileup has several
effects on the detected X-ray events, most importantly that the overall
detected count rate is lowered, and the spectrum is shifted from
lower to higher energies, due to the absence of real lower energy
events and the addition of ``fake'' higher-energy events.

There is also an effect on the ``grade'' of the detected events.  The
grade of an X-ray event is a number that refers to how the charge
cloud from a photon hitting a CCD pixel is spread across neighboring
pixels.  Real X-ray events have characteristic grades that can be used
to distinguish them from unwanted events such as particle background
interactions.  However, when two or more photons are piled up, they
often result in grades that do not match those of individual photons,
and so are thrown away by this filtering, thus decreasing the count
rate.  Taking into account this ``grade migration'' is an important
part of modeling pileup.  For a discussion of these effects in
 detail, see \citet{davi01}.

In this study we deal with pileup in three different ways: short frame
times, excluding events from the center of the PSF (for \xmm), and
pileup modeling (for \chandra).  All five observations have short
frame times $<$75 ms, which mitigate pileup by
reducing the likelihood that multiple photons will be coincident in
one time resolution element.  

For the \xmm\ data, we tested for pileup using the SAS tool
\textit{epatplot}, which compares the distribution of events of
different grades with that expected for no pileup.  When events from
the full PSF were included, these distributions differed
significantly, indicating pileup.  However, because count rates (and
thus pileup effects) are greatest at the center of the PSF, we
found that by excluding events from the central 1/3 radius of the
source region, we could eliminate any significant pileup effects.

For the \chandra\ data we could not simply exclude the inner
extraction region.  Because the PSF is smaller than for the \xmm\, and
because the use of CC mode concentrates all the events along one
dimension, most source events are concentrated in a few pixels.
Therefore, we fit the data in the ISIS package using the pileup
kernel, which models and corrects for pileup effects in spectral
fitting \citep{davi01}.  This kernel includes a grade migration
parameter $\alpha$ which can be allowed to float to best characterize
the effect that pileup has on the grade selection of the observation.
On average 17\% of the photons in the ACIS spectra are piled up.  This
level of pileup can be corrected using the ISIS kernel while measuring
the spectral parameters.

\subsection{Pulse-phase spectroscopy}
Pulse phase-resolved spectra were extracted in the range 0.6--10 keV.
For the \chandra\ data we excluded the energy range 1.7--2.8 keV in
our spectral fits \citep[as in][]{neil04}, to avoid instrumental effects near the Au and Si
edges.  For both \xmm\ and \chandra, we cleaned the data for flaring
episodes before spectral extraction.

\begin{figure}
\epsscale{1.}
\plotone{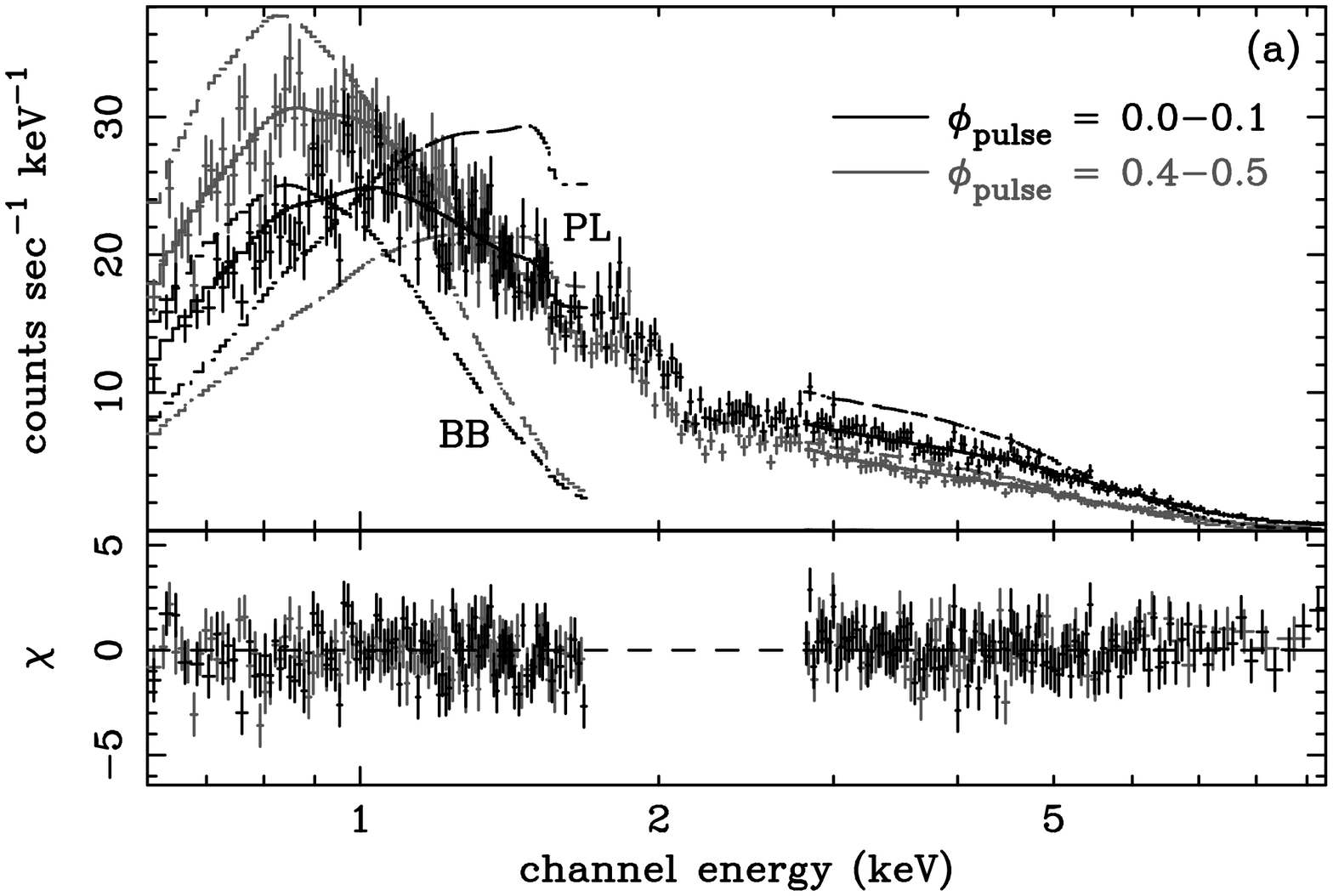}
\plotone{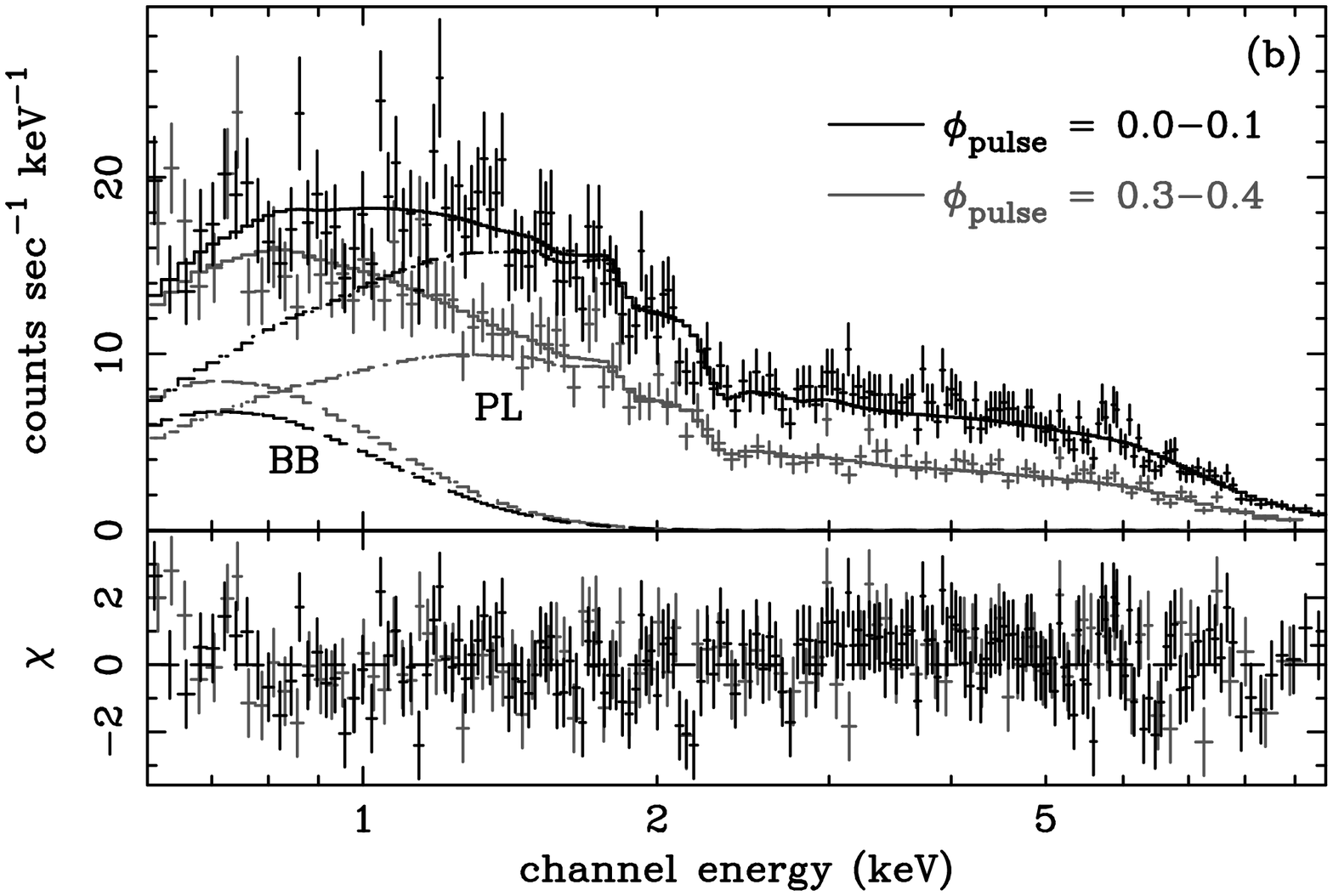}
\plotone{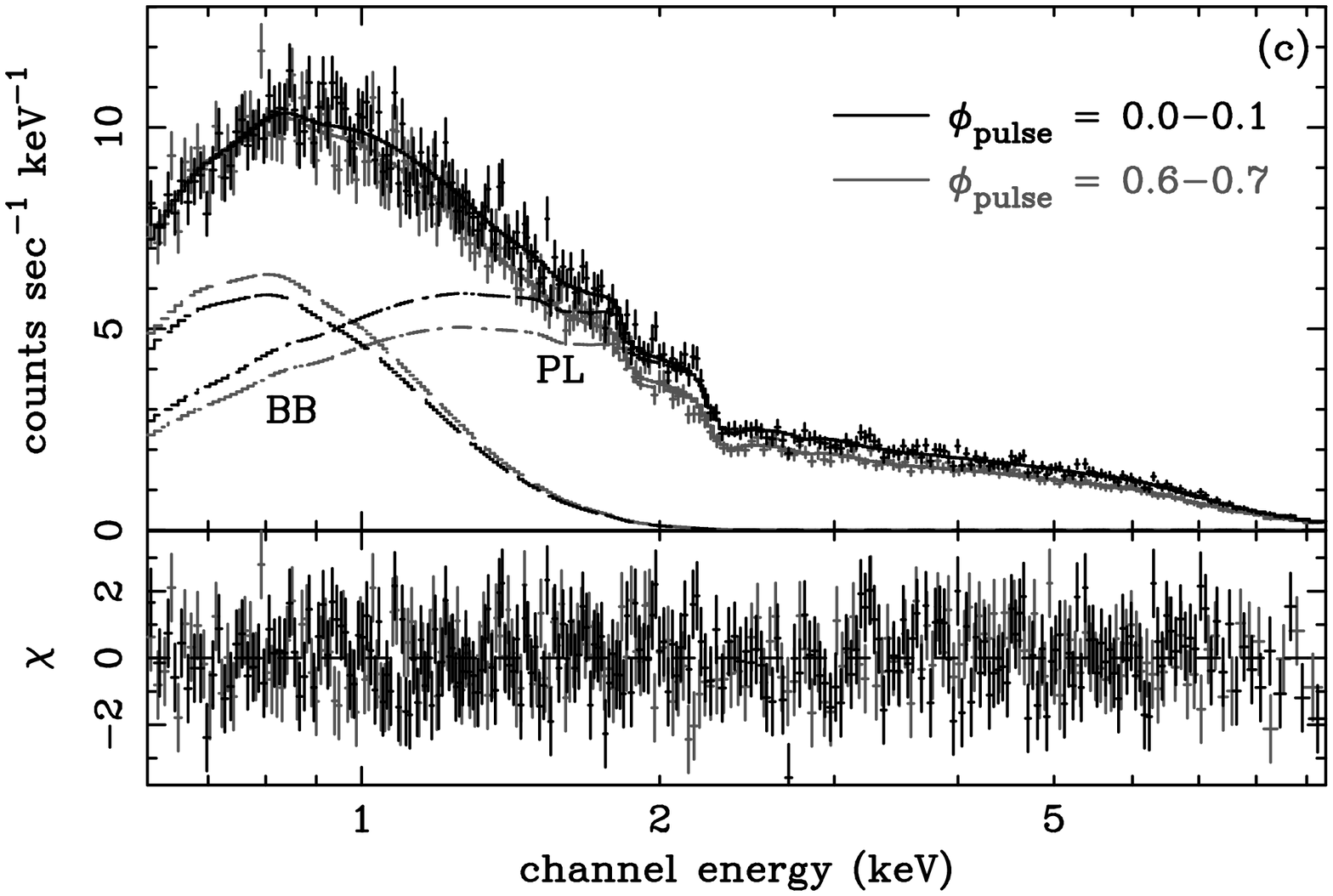}
\caption{Pulse phase-resolved spectra for SMC X-1, for
(a) C103 (b) X101, (c) X201.  For each observation two spectra are
shown, from pulse phases that show substantial differences in the
spectral parameters.  The individual blackbody (BB) and power law (PL)
components are also shown.  In the \chandra\ data, these
individual components do not add to the total, because of the effect
of the pileup model.  Also note that the energy range 1.7--2.8 keV was
excluded from the \chandra\ fits.}
\end{figure}

The best fit spectral models for the \chandra\ and \xmm\ observations
consisted of a power law ($\Gamma \sim 0.9$) plus blackbody $kT_{\rm
BB} \sim 0.18$, both modified by neutral absorption.  We also included
a high-energy exponential cutoff on the power law, as has been seen in \ginga\ and
\asca\ observations \citep{woo95, paul02}.  X201 was the only
observation with sufficient high-energy counts to constrain this
cutoff, so for the other observations we fixed the parameters to the
phase-averaged value for X201 ($E_{\rm cut}=6.1$ keV and $E_{\rm
fold}=6.8$ keV).  The models can be written as:
\begin{description}
\item[\chandra:] $f(E)=K_{\rm pile} \otimes {e^{-\sigma (E) N_{\rm
H}}[f_{\rm BB}(E)+f_{PL}(E)f_{\rm cut}(E)]}$
\item[\xmm:] $f(E)=e^{-\sigma (E) N_{\rm
H}}[f_{\rm BB}(E)+f_{PL}(E)f_{\rm cut}(E)]$
\end{description}
where $f_{\rm BB}$, $f_{PL}$, $f_{\rm cut}$, are the standard XSPEC
blackbody, power law, and high-$E$ cutoff models.  $K_{\rm pile}$ is
the ISIS pileup kernel.

In general, for phase-averaged spectral parameters we use the results
of fits by \citet{neil04}.  However for X101 we performed a new
phase-averaged fit, because we have revised the time interval over
which the spectrum was extracted (see \S\ 2).  We fixed the high-energy
cutoff as above, giving $N_{\rm H}=1.3\times10^{21}$ cm$^{-2}$,
$kT_{\rm BB}=0.17$ keV, and $\Gamma=0.79$, with XSPEC normalizations
$K_{\rm BB}=8.8\times10^{-4}$ (in units of $8.3\times10^{-8}$ \flux)
and $K_{\rm PL}=3.5\times10^{-2}$ (photons keV$^{-1}$
cm$^{-2}$ s${-1}$ at 1 keV).

To extract phase-resolved spectra, we first corrected the photon
arrival times for the motion of the Earth and the observatory, as well
as the orbital motion of SMC X-1 using the ephemeris found by \citet{wojd98}. Using the pulse
periods found by \citet{neil04}, we extracted spectra in 10 bins of
$\phi_{\rm pulse}$ for each of the observations.  We fit the spectra
in ISIS for \chandra\ and XSPEC for \xmm, using the models given
above.  For the \chandra\ fits, we fixed the grade migration parameter
$\alpha$ at its phase-averaged value.

\begin{figure}
\plotone{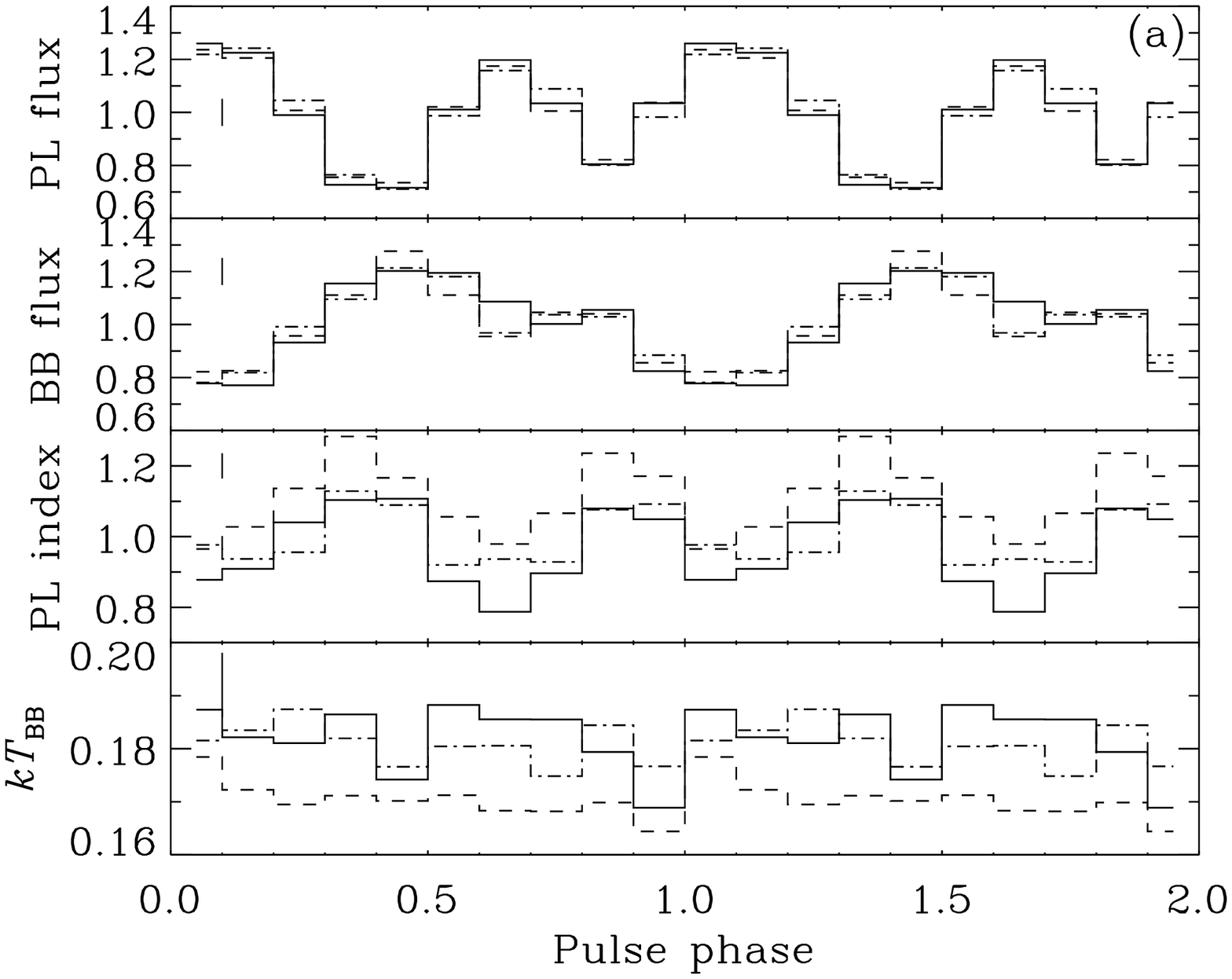}
\plotone{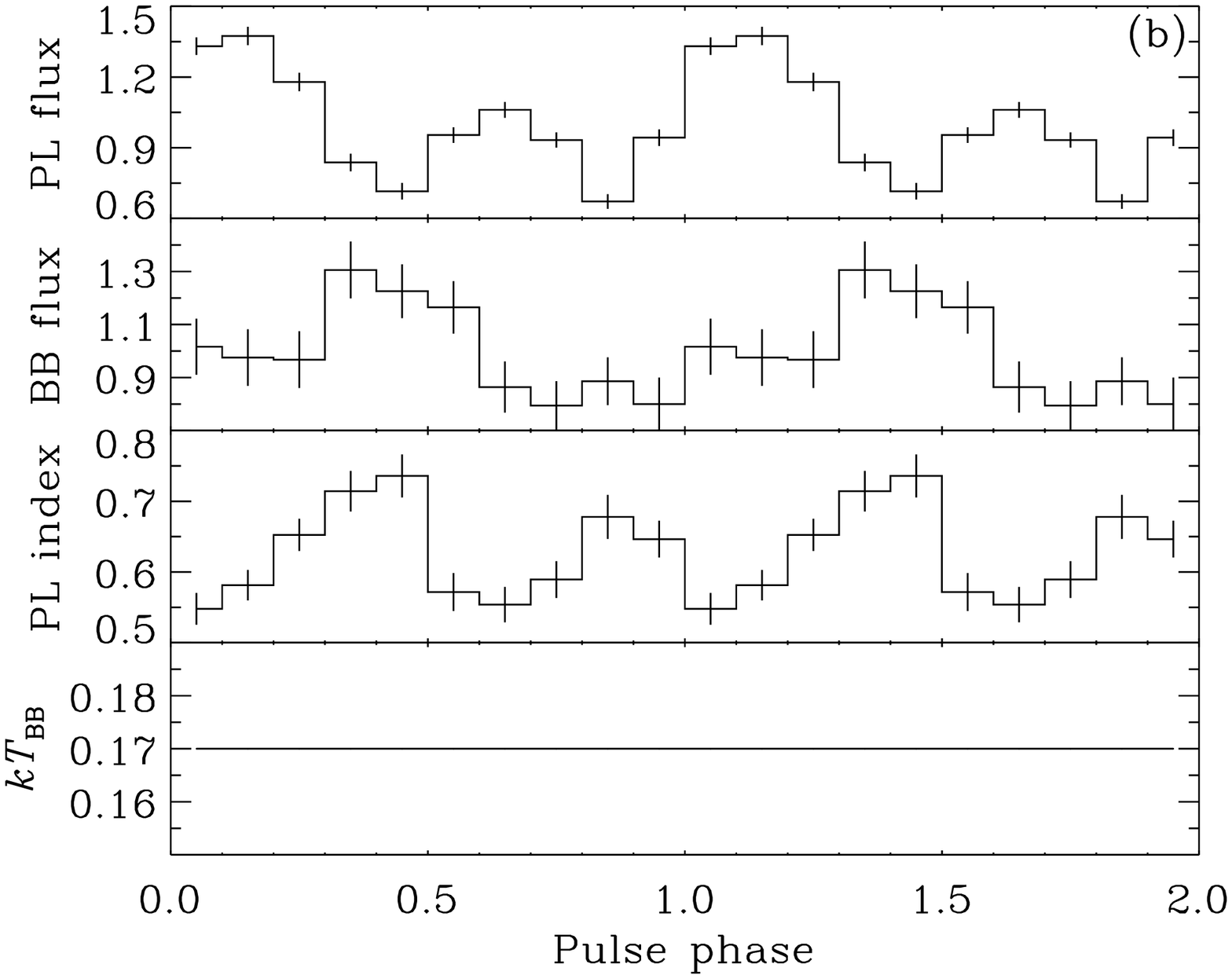}
\plotone{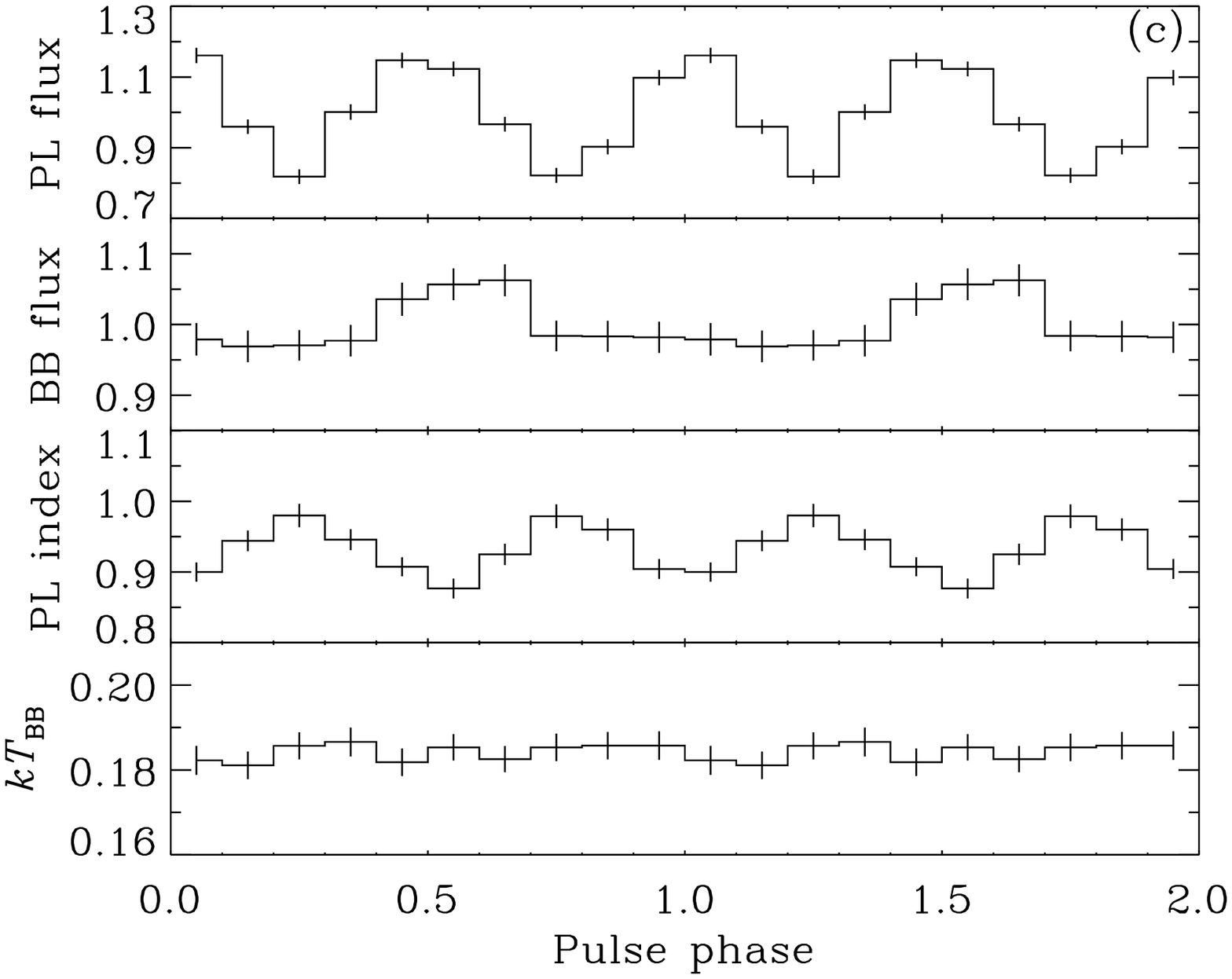}
\caption{Pulse phase variation of spectral parameters for SMC X-1, for
(a) \chandra: solid: C102, dashed: C103, dot-dashed: C104 (b) X101,
(c) X201.  Shown are the relative fluxes from the power law and
blackbody components, the power law index $\Gamma$, and the blackbody
temperature $kT_{\rm BB}$ in keV (note that the temperature was fixed in the
X101 fits).  Two pulses are shown for clarity.  Errors are 1$\sigma$,
and typical error bars for \chandra\ are given to the left of the
plots.  The average fluxes in the power law and blackbody components
are given in Table 1.}
\end{figure}

For each fit we set the initial parameters to the phase-averaged
values, and fixed the $N_{\rm H}$ and high energy cutoff while
allowing the other parameters to vary.  For X101, the limited
statistics required that we also fix the blackbody temperature,
$kT_{\rm BB}$, in order to measure the variation in the soft
component.  Sample phase-resolved spectra for C103, X101, and X201 are
shown in Fig. 4, for which changes in shape between spectra at
different $\phi_{\rm pulse}$ are clearly visible.  The phase-resolved
fits have $\chi^2_{\nu}$ values of 0.8--1.0 for \chandra, 1.0--1.3 for
X101, and 0.9--1.2 for X201.

We found that the best-fit spectral parameters vary significantly with
$\phi_{\rm pulse}$. For each phase we calculated the 0.5--10 keV
unabsorbed flux in both the blackbody and (cutoff) power law
components. The $\phi_{\rm pulse}$ variation of the fluxes, along
with $\Gamma$ and $kT_{\rm BB}$, are shown in Fig. 5.  We find that:
\begin{enumerate}
\im At every epoch, the power law flux shows the familiar double
peaked profile, similar to the hard pulse profile observed many times
previously.  The phase offset and relative intensities of the two
peaks vary slightly between epochs.

\begin{figure}
\plotone{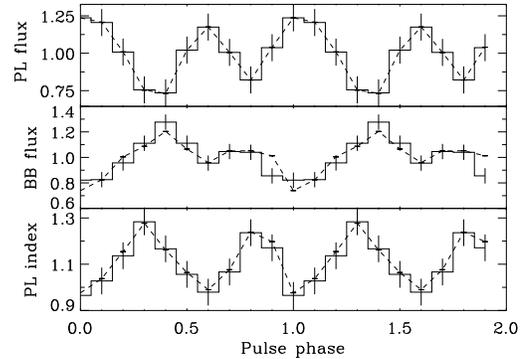}
\caption{Pulse-phase variation in spectral parameters for C103, using a
blackbody (histogram) and a thermal bremsstrahlung (dashed line) soft
component model.  Note that the hard and soft flux profiles do not
differ significantly for the two models.}
\end{figure}

\im The power law photon index varies significantly, inversely with the
power law flux.  The spectrum gets harder as the power law flux
increases, and softer as the flux decreases.  

\im For all observations, the blackbody flux shows a single main peak,
and in the \chandra\ observations there is a shoulder after the main
peak.

\im The shapes and relative phases of the blackbody pulses
change with time.  However the three \chandra\ observations (at $\phi_{\rm
orb}=0.19$, 0.49, and 0.74) have very similar soft pulses. This
implies that the changes in the soft profiles are not correlated with
orbital motion of the source, and occur on timescales longer than a few
days.  
\end{enumerate}

To test whether the observed variation is dependent on the particular
choice of soft component model, we repeated the pulse-phase
spectroscopy using a thermal bremsstrahlung (TB) model in place of the
blackbody.  This fits are acceptable for $kT_{TB}\sim0.3$ keV, with
very similar $\chi^2$ to the blackbody fits.  This is not a physical
model, because a large, diffuse thermal gas cloud could not produce
the luminous, pulsing soft emission observed \citep{hick04}, but it
allows us to check that the measurement of variation in the soft
component is robust.  The soft and hard component flux profiles for
this model for one observation (C103) are given in Fig. 6, and are
compared to those for the blackbody model.  The shapes of the profiles
differ by $\lesssim0.5\sigma$, for the two models, except for the soft
profile in C102, for which they differ by $\simeq1.2\sigma$.  Since
these differences are not highly significant, we conclude that the
hard and soft flux profiles are essentially independent of the model
chosen for the soft component.

\section{Photon index variation}
The observed changes in the power law index with pulse phase have not
been previously reported for SMC X-1; the \asca\ study of
\citet{paul02} found that $\Gamma \simeq 1.0$ for four different pulse
phases.  We have investigated the possibility that this variation may
not be real, but an instrumental effect due to pileup.  This is
possible in principle because pileup increases with flux and serves to
lower soft count rates and increase hard count rates, any residual
pileup for which we have not accounted could cause a ``hardening'' of
the spectrum as flux increases.  However after detailed testing we
conclude that the variation is not due to pileup effects, because:
\begin{enumerate}
\im Modeling of pileup in the \chandra\ data using the ISIS
pileup kernel \citep{davi01} shows that the $\sim$50\% changes in the
flux that we observe can only change the power law slope by 0.01,
which is much smaller than the observed variation in the \chandra\
spectrum.  

\im Variation in $\Gamma$ is also observed in the \xmm\ observations,
for which piled-up events have been removed.  If we use all the events
(including piled-up ones) in the X201 analysis, we find that the {\it
average} value of $\Gamma$ is flatter by $\sim$0.1 due to pileup
effects, but the variation in $\Gamma$ of $\pm$0.5 remains the same.
\end{enumerate}
Therefore the pulse variation in the power law index appears to be
independent of any pileup.  We treat this effect as real and
consider physical interpretations.  

Power-law index variation with pulse phase has been observed in a
number of X-ray pulsars.  We searched the sample of bright pulsars in
Table 2 of \citet{hick04} for those with existing pulse-phase
spectroscopy.  We find that many sources
show variations in $\Gamma$, and these tend to be correlated with the
pulse intensity.  Some sources, such as LMC X-4 \citep{woo96}, Cen X-3 \citep{burd00}, Vela
X-1 \citep{kret97}, and \fouru\ \citep{prav79}, become harder with
increasing intensity, as we have observed for SMC X-1.  However, Her X-1 \citep{rams02},
V0332+53 \citep{unge92}, and 4U 1538--52 \citep{clar94} become {\it softer} with increasing flux.
\xtej\ \citep{yoko00}, \rxj\ \citep{kohn00} , and GX 301--2
\citep{endo00} show no variation in $\Gamma$.

Because of this variety of behaviors, it is difficult to present a
universal physical picture for the pulse-phase changes in the power
law spectra of X-ray pulsars.  In the standard picture of spectral
formation in luminous pulsar beams, the accretion column is stopped by
a radiatively driven shock.  If photons in the brightest part of the
beam are generated in strongest shock, they will have been scattered
with larger Compton $y$ parameter and so preferentially emerge at
higher energies.  This would make the center of the pulse harder and
the outside softer, as observed for SMC X-1.  A mechanism for a
softening with pulse flux is not immediately clear.  We would like to
stress, however, that variation in $\Gamma$ with pulse phase is very
common, and so must be addressed by successful models of X-ray pulsar
spectral formation.

\section{Reprocessing by a warped disk}
The soft component pulses observed are roughly consistent
with the picture argued by \citet{hick04} and \citet{neil04} that
double-peaked power law pulses come from near the neutron star's
accreting polar caps, while the soft pulses come from a much larger
radius, at the irradiated inner region of the accretion disk.

We seek to test this possibility by comparing the observed hard and
soft pulse profiles to a relatively simple model, in which a warped,
twisted inner accretion disk is illuminated by a rotating X-ray pulsar
beam.

\begin{figure}
\epsscale{1.0}
\plotone{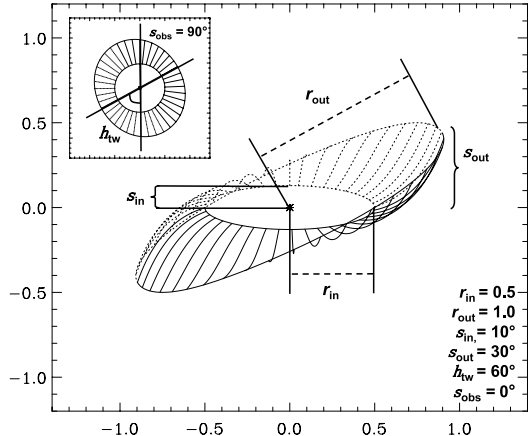}
\caption{Diagram of the twisted disk model, showing the relevant
parameters.  The disk is viewed at elevation $\theta_{\rm
obs}=0^{\circ}$. {\it Inset}: the same disk viewed at $\theta_{\rm
obs}=90^{\circ}$, showing the disk twist angle $\phi_{\rm tw}$.}
\end{figure}

\subsection{Illuminated disk model}
Our model does not describe the entire disk, but only the region close to the
magnetosphere where the bright reprocessed component must be emitted.
The disk model consists of a series of concentric circles with varying tilt
and twist relative to each other with distance from the center (see
Fig. 7.  A similar shape has been inferred for the large-scale
structure of the disk in Her X-1, from changes in the pulse profiles
with superorbital phase \citep{scot00,leah02}.  Warped
and twisted structures have also resulted from numerical simulations of
warped disks \citep{wije99} and of gas-magnetosphere interaction
\citep{roma03}.

\begin{deluxetable*}{lrrrr}
%\tabletypesize{\scriptsize}
\tablecaption{Parameters for disk models}
\tablewidth{0pt}
\tablehead{ \colhead{} & \multicolumn{2}{c}{Model ranges} & 
\multicolumn{2}{c}{Best fit} \\
\colhead{Parameter} & 
\colhead{Pencil beam} &
\colhead{Fan beam} &
\colhead{Pencil beam} &
\colhead{Fan beam}}
\startdata
$L_X$ (\ergs) & $3 \times 10^{38}$ & $3 \times 10^{38}$ & $3 \times
10^{38}$ &  $3 \times 10^{38}$ \\
$R_{\rm in}$ ($10^8$ m) & 0.8 & 0.8 & 0.8 & 0.8\\
$R_{\rm out}$ ($10^8$ m) & 1    & 1 & 1 & 1 \\
Outer angle $\tout$ (deg)  &  30 & 30 & 30 & 30 \\
Inner angle $\tin$  (deg)       &  10 & 10 & 10 & 10 \\
Twist angle $\phi_{\rm tw}$  (deg)      & -90, 90, 135 &
 90 & 90 & 90 \\
$\theta_{\rm b1}$  (deg)  & 35, 45 & 0, 45 & 35 & 45 \\
$\theta_{\rm b2}$  (deg)  & -10, 0, 10, 20 & 0, 45 & -10, 10\tnm{a} &
0 \\
$\phi_{\rm b1}$  (deg)  & 0 & 0 & 0, & 0 \\
$\phi_{\rm b2}$  (deg)  & 210, 225, 240 & 180, 210, 225 & 225,
210\tnm{a} & 180, 210\tnm{a} \\
Beam half-width $\sigma_{\rm b}$ (deg) & 30 & 15 & 30 & 15\\
Fan opening angle $\theta_{\rm fan}$  (deg) & 0 & 45, 60, 90 & 0 & 60 \\
Observer elevation $\theta_{\rm obs}$    (deg)    &  20 & 20 & 20 & 20
\enddata
\tnt{a}{These beam parameters are required to fit the X201 pulse profiles.}
\end{deluxetable*}

In our calculations we place this surface (we will refer to it simply
as the ``disk'') at roughly the location of the magnetosphere
($\sim$$10^8$ cm), and test if precession of such a surface around the
disk axis can create changing soft-component profiles
similar to those observed.  The shape of the surface is determined by
the radii and tilt angles of the inner and outer circle ($\rin$,
$\rout$, $\tin$ and $\tout$), and the offset angle or ``twist''
between them ($\phi_{\rm tw}$).  Letting $\theta_{\rm d}$ be the angle
of the disk above the rotation plane at any point ($r,\phi$), we use
the simple approximation:
%\begin{equation}
%
$$\theta_{\rm d}(r,\phi)=-(\tin+(\tout-\tin)\frac{r-\rin}{\rout-\rin})
\sin{(\phi-\phi_{\rm tw})},$$
%
%\end{equation}
which is valid for $\theta \lesssim 30^{\circ}$.

It is also necessary to define a beam pattern from the neutron star.
Complex beam shapes have been proposed for some X-ray pulsars; for
example Her X-1's emission has been modeled as a central pencil beam
surrounded by a fan beam \citep{blum00}, or even a reverse-pointing
fan beam, emitted from above the neutron star's surface
\citep{scot00}.  However, since we are modeling coarse pulse
profiles with only 10 bins of pulse phase, it is impossible to
constrain such complex models, so we use a simple
beam shape.  

The beam is described by two two-dimensional gaussians, superposed
 on an isotropic component which is required since the hard pulse
 fraction for SMC X-1 is only $\sim$50\%.  We define the beam pattern in
spherical coordinates, with the poles aligned with the rotational axis
and $\theta=0$ at the equator. We set the neutron star's rotation to
be parallel to the disk axis. Each gaussian is specified by its width
($\sigma_{\rm b}$), its angle from the rotational plane ($\theta_{\rm
b}$), and its azimuthal angle ($\phi_{\rm b}$).  We also define a fan
opening angle ($\theta_{\rm fan}$), which we set to zero to model
``pencil'' beams, and to $\geq 45$\dgr\ to model ``fan'' beams.

We set the peak intensity of each gaussian equal to 3 times the
intensity of the isotropic component.  Note that this normalization is somewhat
arbitrary; when we fit the output model profiles to the data
we allow the intensity of the pulses to vary.  The beam pattern is
given by:
$$F_{\rm b}(\theta, \phi)=1+3 e^{-(\alpha_1-\theta_{\rm
fan1})^2/2\sigma_1^2}+3 e^{-(\alpha_2-\theta_{\rm fan2})^2/2\sigma_2^2}$$
where $\alpha_{1,2}$ is the angular distance between
($\theta,\phi$), and ($\theta_{\rm b1,2},\phi_{\rm b1,2}$).
This beam pattern is normalized to a total hard emission luminosity of
$L_{\rm X}=3 \times 10^{38}$ \ergs.

This beam pattern is then swept around the disk.  We take the disk to
be opaque and calculate the luminosity absorbed by each patch of the disk
surface.  We assume that all this energy is re-radiated as a blackbody
spectrum, with temperature
	$$T=(\textrm{d}L/(4 \pi \sigma \textrm{d} A))^{1/4}.$$  
Only those regions of the disk that are directly visible from the
neutron star are illuminated.  Finally, we calculate which of the
illuminated regions are visible to the observer, and are not blocked
by parts of the disk in front.  

We assume that the heated disk immediately re-radiates its thermal
energy, so that there is no time lag between illumination by the X-ray
beam and emission from the disk.  This assumption requires that both
the light-crossing time for the system and the cooling time for the
heated disk are much shorter than the pulse period.  In our model the
illuminated region is at $R\sim10^8$ cm, so the light crossing time is
$\sim$10 ms.

A rough estimate for the cooling timescale is given by \citet{endo00},
and is simply the total thermal energy of the heated disk divided by
its luminosity.  The disk's thermal energy is the volume of the
heated region times its thermal energy density.  If we treat the disk
as a partial spherical shell presenting solid angle $\Omega$ to the
neutron star, we have
$$E_{\rm BB}=\Omega 4 \pi R_{\rm BB}^2 d \left ( \frac{3}{2} n kT_{\rm
BB} \right ),$$ where $d$ is the penetration depth of hard X-rays into
the heated region.  We assume that most X-rays penetrate no more than
a few Compton depths, so that $nd \sim 3/\sigma_{\rm T} \sim 10^{25}$
cm$^{-2}$.  For $R_{\rm BB} = 10^8$ cm, $\Omega/4 \pi\sim 0.1$,
and $kT_{\rm BB} = 0.18$ keV, we have $E_{\rm BB} \sim 10^{32}$ erg.
The cooling timescale is simply $t_{\rm cool}=E_{\rm BB}/L_{\rm
soft}$, which for $L_{\rm soft} \sim 10^{37}$ \ergs\ gives $t_{\rm
cool} \sim 10^{-5}$ s.

Since both these timescales are much shorter than the pulse period, we
can treat the reprocessing as instantaneous.  For each
neutron star rotation phase, we calculate the total luminosity visible
to the observer for the hard emission from the neutron star,
appropriate to the observer's view of the given beam pattern.  We also
sum the observable reprocessed emission from the disk.  We thus obtain
``hard'' and ``soft'' model pulse profiles that we can compare to observations.

\begin{figure}
\epsscale{1.}
\plotone{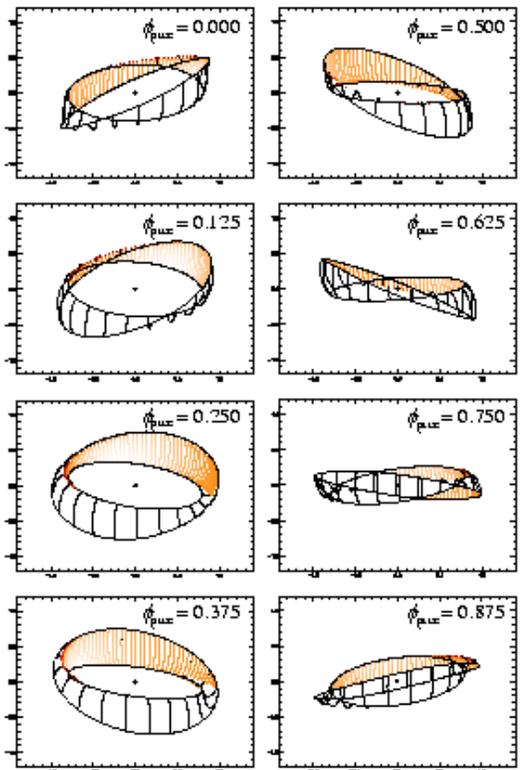}
\caption{Model twisted, tilted accretion disk as used in our
reprocessing calculations.  Shown is the model which allows the best
fit to the observed (see Table 2).  The shaded regions show areas on
the disk which are illuminated by the central source and are visible
to the observer. Lengths are in $10^8$ cm.}
\end{figure}

\subsection{Model output}
We have created a number of illuminated disk models, varying the disk
shape ($\tin,\tout,\phi_{\rm tw}$), the beam pattern ($\theta_{\rm
b1},\theta_{\rm b2},\phi_{\rm b2},\theta_{\rm fan}$), and the observer's elevation
angle $\theta_{\rm obs}$.  Because the model is
computationally intensive, we did not perform a systematic study of
the whole range of possible configurations.  We initially ran the
model for a broad variety of disk shapes and beam patterns, then limited our
systematic study to those parameters that could roughly reproduce the
observed properties of SMC X-1.  These parameter ranges are given in
Table 2.  We set the outer disk to be inclined at 30\dgr, which is
consistent with disk inclination estimates of 25\dgr--58\dgr
\citep{luto04}, and set the inner disk at a somewhat smaller
inclination of 10\dgr.  This disk shape results in 20--30\% of the
neutron star luminosity being reprocessed. We set the
elevation angle of the observer, with respect to the neutron star's
rotation plane, to be $\theta_{\rm obs}=20$\dgr.   If the
neutron star spins in the plane of its orbit, this corresponds to an
orbital inclination $i=70$\dgr.  This is consistent
with the estimated range of $i\simeq65$--70\dgr\ from optical studies,
assuming approximately Roche geometry for the companion star \citep{reyn93,hutc77}. To obtain the
observed blackbody temperature of $\sim$0.18 keV, we placed the
reprocessing region between $r_{\rm in}=0.8\times 10^8$ cm and $r_{\rm
out}=1\times 10^8$ cm.

The model disks consisted of a grid on $(r,\phi)$ of 100 elements in
each dimension.  For each disk shape we calculated the emission at 30
separate pulse phases and 8 disk precession phases.  The observer's views of the disk at
several precession phases are shown in Fig. 8.  In the upper left
panel of Fig. 8, the neutron star has just emerged from behind the
precessing disk, which corresponds to the start of the high
superorbital intensity state for which $\phi_{\rm SO}=0$.  Therefore
we set $\phi_{\rm prec}=0$ for this model disk orientation and define the
other precession phases accordingly.  For each model beam and
$\phi_{\rm prec}$, we calculated the observed pulse profiles for the
direct emission from the neutron star (the hard profile) and the
reprocessed emission from the disk surface (the soft profile).

\subsection{Pencil beam models}
We begin by examining pencil beam shapes, for which $\theta_{\rm
fan}=0$.  To reproduce the hard pulse profiles observed, we found that
the two beams could not be anitipodal, that is $\theta_{\rm b1} \ne
\theta_{\rm b2}$ and $\phi_{\rm b2} \ne 180$\dgr (with $\phi_{\rm
b1}= 0$\dgr).  The observed hard pulse peaks are not exactly
180\dgr\ out of phase, so we varied $\phi_{\rm b2}$ between 210\dgr\ and
240\dgr.  To produce the relative heights of the two hard pulses, we
set $\theta_{\rm b1}$ to 35\dgr\ or 45\dgr\ and varied $\theta_{\rm
b2}$ between -10\dgr\ and 20\dgr.  

While these represent only a small subset of the possible
disk/beam configurations, they allow us to test whether, in principle,
reprocessing by a precessing disk can reproduce the observed pulse
profiles.   The model results for pencil beams show that:

\begin{enumerate}
\im Most disk configurations display some kind of variation in the shape
and phase of the pulses in the soft, reprocessed component.

\im The relative shapes of the hard and soft pulses vary
most strongly with the shape of the illuminating beams, and less
strongly with the detailed shape of the disk.

\end{enumerate}

For our array of models, we performed $\chi^2$ fits of the model
pulses to the observed power law and blackbody flux profiles (shown in
Fig. 5) for each observation.  We allowed the normalization of the
calculated pulses to vary, in order to account for any differences in
pulse fraction between the observed pulses and our model.

For each model we first fit the hard profiles to the observed
power-law profiles, allowing the scale and the phase offset of
the model profiles to vary.  This achieved a reasonable fit
($\chi^2_\nu \lesssim 2$) for many pencil-beam models.  We next
attempted to fit the blackbody pulses by varying only the amplitude of
the model profiles; the phase was fixed to that from the hard profile
fit.  For most model parameters, the shape and phase offset of the
model soft pulses do not match those in the data, and no good
fit is achieved.

\begin{figure*}
\epsscale{1.0}
\plotone{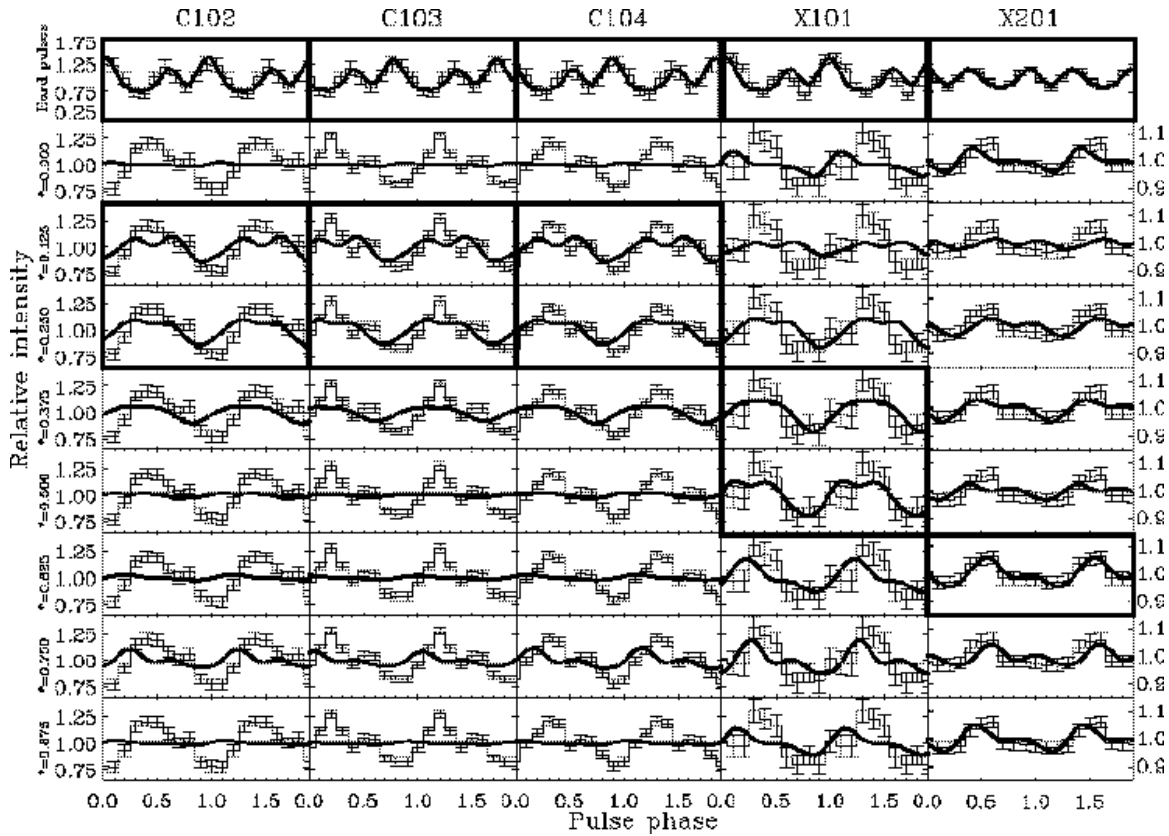}
\caption{Hard and soft pulse profiles output by the disk reprocessing
model, fitted to the observed profiles for the five observations.  The
disk and beam parameters are given in Table 2.  Each column shows
observed spectral component pulses (as in Fig. 5) from one of the five
observations.  The top row shows the power law profiles fitted by the
pulses from the neutron star, and the lower rows show blackbody
profiles fitted by the model reprocessed pulses.  Reprocessing for 8
disk precession phases $\phi_{\rm prec}$ are shown.  Note that the
observed soft pulses vary in phase and shape with $\phi_{\rm prec}$.
Highlighted in bold are the profiles that best fit for the observed
power-law pulses (top) and reprocessed pulses (below).  This disk and
beam configuration fits the \chandra\ data best at $\phi_{\rm
prec}=0.125$--0.25 and X101 best at $\phi_{\rm prec}=0.375$--0.5.  The
X201 pulses require a slightly different beam pattern (see Table 2)
and fit the data well for $\phi_{\rm prec}= 0.625$.  Note that the
X201 soft pulses are weaker than the other four, and are shown using a
different scale.}
\end{figure*}

\begin{figure*}[p]
\plotone{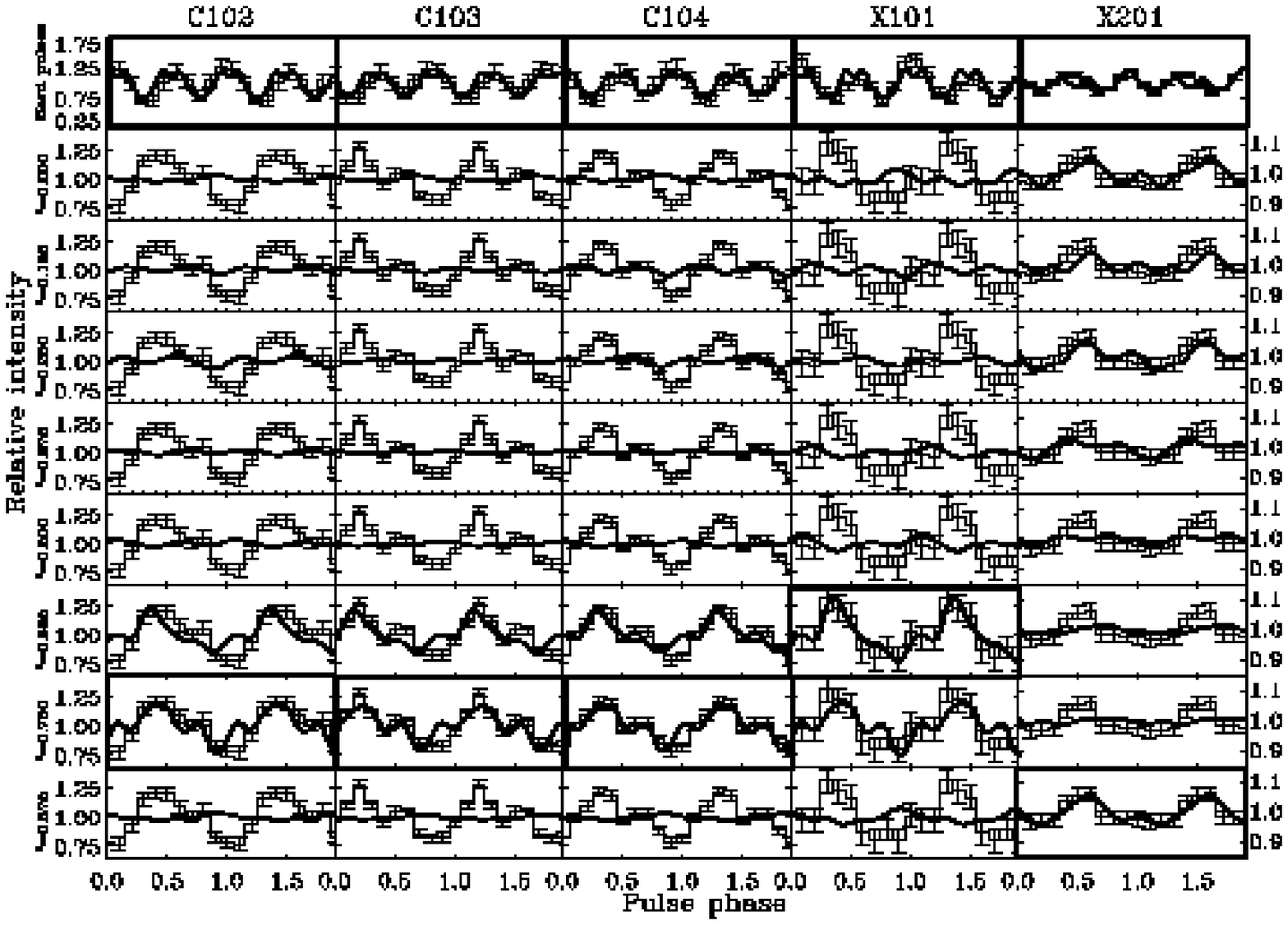}
\caption{Same as Fig. 8, for a fan rather than a pencil beam shape.
The profiles correspond to the fan beam configuration that best fit the
observations, with parameters given in Table 2.}
\end{figure*}

However, for one particular disk shape (Fig. 8, parameters given in
Table 2), we have achieved reasonable fits to the hard and soft
profiles for all five observations (see Fig. 9). Between the
observations we needed to change only: (1) the precession phase of the
disk, and (2) for X201, the beam orientation ($\theta_{\rm b2}$ and
$\phi_{\rm b2}$). The reduced $\chi^2$ values for the individual fits
are given in Table 3.  Note that for the \chandra\ soft pulses, even
the best fits have $\chi^2_\nu > 5$ due to the complex shapes and
small errors on the observed pulses.  Qualitatively, however, the soft
profiles have roughly the right features and phase offset, so we
consider the fits satisfactory.  The values in Table 3 show that for
each observation, there is a clear minimum in the $\chi^2$ values as a
function of $\phi_{\rm prec}$.  This occurs because the reprocessed
pulse profiles vary as the disk precesses and thus fit the observations
for only specific disk orientations.

 The soft component pulses are fit best for $\phi_{\rm
prec}=0.125$--0.25 for \chandra, 0.375--0.5 for X101, and 0.75 for
X201 (corresponding to the bold boxes in Fig. 9).  For \chandra\ and
X101 these are consistent with the $\phi_{\rm SO}$ estimates of 0.16
and 0.37.  There is not good agreement for X201, for which $\phi_{\rm
SO}=0.42$.  However we note that X201 is also fit reasonably well for
other phases including $\phi_{\rm prec}=0.375$, which would be
consistent with its observed superorbital phase.

\subsection{Fan beam models}
We have also used our model to examine reprocessing of a fan-shaped
pulsar beam, for $\theta_{\rm fan}\geq45$\dgr.  For the pencil beam we
found that disk parameters do not strongly affect the reprocessed
pulse profiles, so for the fan beam we fixed the disk parameters at
the best-fit values for the pencil-beam case.  We then varied the beam
parameters with the values given in Table 2, setting the fan opening
angles equal for both beams ($\theta_{\rm fan1}=\theta_{\rm
fan2}=\theta_{\rm fan}$).  In general, the fan beam is less successful
than the pencil beam in reproducing the observed hard pulses, because
in most cases the fan beam produces three or four peaks in the hard
profile as both edges of each fan sweep past the observer.

However, in the cases where we view the edge of each fan, or when
there is only one fan ($\theta_{\rm fan}=90$\dgr), we see only two
hard pulses.  For one beam configuration, we find fits that are roughly
comparable to those for our best-fit pencil beam case (see Fig. 10 and
Table 3).  As for the pencil beam case, the fits to X201 require a
slightly different beam configuration ($\phi_{\rm b2}=210$\dgr\ rather
than 180\dgr).

Again we find that the best model fits occur for different $\phi_{\rm
prec}$ for different observations, but in this case the best fits are
for models with late disk precession phases ($\phi_{\rm prec}=0.625$
for X101, 0.75 for \chandra, and 0.875 for X201).  This is not
consistent with these observed $\phi_{\rm SO}$, which range from 0.16
to 0.42.  In fact late precession phases correspond to the low state
of the superorbital cycle, when the disk would be covering the neutron
star (Fig. 8).  Therefore we do not expect that these model parameters
give a realistic description of the system.  However, this result does
suggest that in principle a fan beam, like a pencil beam, might
produce the changing soft profiles observed.

\section{Discussion}
We have constructed a model of reprocessed pulses from a warped,
precessing X-ray pulsar disk, and have succeeded in roughly
reproducing the changing power law and blackbody pulse profiles
observed in SMC X-1.  This suggests that reprocessing by the accretion
disk is a plausible explanation for the observed X-ray pulses in SMC
X-1, and is supported by the fact that other possible origins of the
soft component can be ruled out on physical grounds \citep{hick04}.
Because our model has a number of adjustable parameters, we cannot place
realistic constraints on the disk and beam geometries.  However, we
can conclude that the long-term evolution of the SMC X-1 spectral
behavior is consistent with precession of an illuminated inner disk.

It remains unclear whether the long-term behavior of the power law and
blackbody pulse profiles for SMC X-1 is in fact correlated with the
superorbital period.  In an \xmm\ study of Her X-1, \citet{zane04}
found that changes in the phase offset between the hard and soft
pulses may vary monotonically with superorbital period, and so may be
directly related to the disk precession.  We are unable to say this
for SMC X-1, because our determination of the superorbital phases is
uncertain and because we only have data from a few epochs.  Additional
data would be very useful for solving this problem.  Of the existing
data taken since the launch of \rxte\ (which is required to determine
$\phi_{\rm SO}$) there is only one high-state observation, with
\bepposax, 2 March 1997 for which pulse phase-resolved spectroscopy
has been performed \citep{naik04a}.  This observation was taken at
$\phi_{\rm SO}\sim0.5$, and shows the familiar two-peaked power law
profile and a wide blackbody pulse, offset from the power law pulses,
which is broadly consistent with the observations presented here.
Other studies, such as pulse-phase spectroscopy performed with \asca\
\citep{paul02}, show similar behavior but have no clear determination
for $\phi_{\rm SO}$.

 A future observational program, comprising 6--8 exposures taken at
regular intervals in $\phi_{\rm SO}$ across the high state in a single
superorbital cycle, would be ideal in allowing us to characterize the
long-term evolution by studying gradual changes in the X-ray emission.
In addition, a few exposures at redundant $\phi_{\rm SO}$ in
neighboring superorbital cycles would help test the periodicity of
these variations.   It would be also interesting to perform a more limited study,
similar to the present one, for LMC X-4. If that source shows
long-term variation in the soft profiles which is also consistent with
a disk reprocessing model, this will give support to a consistent
picture of the accretion processes in this group of systems.

\begin{deluxetable}{cccccc}
%\tabletypesize{\scriptsize}
\tablecaption{Reduced $\chi^2$ values for model profile fits\tnm{a}}
\tablewidth{0pt}
\tablehead{ 
\colhead{$\phi_{\rm prec}$}&
\colhead{C102} &
\colhead{C103} &
\colhead{C104} &
\colhead{X101} &
\colhead{X201}}
\startdata
& &  \multicolumn{2}{c}{\it Pencil Beam} \\
Hard\tnm{b} &   \bf{\W0.9}  &  \bf{\W0.8}  &  \bf{\W0.7}  &  \bf{5.0} &   \bf{1.9} \\
%\hline
0.000 & 11.3 &  15.9 &  15.1 &  2.8 &  1.4 \\
0.125 & {\bf \W5.9} & \bf{10.2} &   \bf{\W8.0} &  3.0 &  2.4 \\
0.250 &	{\bf \W6.2} &  \bf{10.3} &   \bf{\W7.9} &  2.2 &  1.8 \\
0.375 &	\W9.2 &  13.4 &  11.6 &  \bf{1.5} &  1.3 \\ 
0.500 &	10.7 &  15.1 &  14.0 &  \bf{1.3} &  2.0 \\
0.625 &	10.9 &  15.4 &  14.1 &  2.1 &  \bf{0.4} \\ 
0.750 &	\W9.5 &  13.9 &  11.7 &  2.2 &  1.5 \\
0.875 &	11.3 &  15.9 &  14.9 &  2.6 &  0.9 \\
& &   \multicolumn{2}{c}{\it{Fan Beam}} \\
Hard\tnm{b}  &  \bf{\W1.7} &  \bf{\W1.2} &  \bf{\W1.1} &  \bf{0.7} &  \bf{5.0} \\
0.000 & 12.0       & 16.9       & 16.7       & 4.0       &      1.1  \\
0.125 & 10.6       & 15.4       & 13.8       & 3.5       &      1.8  \\
0.250 & 10.1       & 15.2       & 13.8       & 3.1       &      1.0  \\
0.375 & 11.8       & 16.5       & 15.4       & 3.6       &      1.8  \\
0.500 & 12.6       & 17.6       & 16.9       & 3.9       &      2.4  \\
0.625 & \W6.1      & \W7.3      & \W6.3      & \bf{0.6}  &      2.4  \\     
0.750 & \bf{\W4.8} & \bf{\W6.2} & \bf{\W5.3} & 1.4       &      2.6  \\
0.875 & 12.8       & 17.5       & 17.5       & 4.0       &      \bf{0.7}  
\enddata
\tnt{a}{See Figs. 9 and 10 for model profiles.  The hard and soft
profile fits have 8 and 9 degrees of freedom, respectively.}
\tnt{b}{For fits to the hard (power law) profiles.}
\end{deluxetable}

For the parameter ranges we have chosen, the pencil beam model
succeeds better than the fan beam in matching the pulse profile data
at the known superorbital phases.  However the beam models are simple
and there is much of parameter space that we have not explored, which
makes any preference of a pencil beam tentative at best.  On
theoretical grounds, the beam shape depends on the details of the
accretion column, with polar cap emission from the neutron star's
surface producing a pencil beam, and emission from the sides of the
accretion column producing a fan beam \citep[e.g.][]{naga89, mesz88}.
 $L_{\rm X}$ for SMC X-1 is close to or at the Eddington limit
($L_{\rm Edd}=1.8\times10^{38}$ \ergs\ for a 1.4 $M_{\sun}$ neutron
star), which suggests that in the accretion column, where the
accretion flow will be most intense, radiation cannot escape out the
top of the column and so a fan beam is required
\citep[e.g.][]{beck98,beck05}.  However, any realistic prescription for
pulsar beams involves a detailed description of the accretion flow as
well as effects such as gravitational light-bending \citep{mesz88,
leah03}. These can create complex beam patterns that may involve both
pencil and fan shapes, and are beyond the scope of our simple model.

While the present work cannot place strong constraints on the beam
geometry, we note that our best-fit pulsar beam for both pencil and
fan shapes is highly inclined with respect to the neutron star's
rotation, and one beam is offset from the antipodal position of the
other beam.  This may suggest that the magnetic field is inclined from
the rotation axis and is not exactly dipole.  Distorted dipole fields
have also been suggested for Her X-1 \citep{blum00} and Cen X-3
\citep{krau96} from detailed analysis of pulse shapes, and this
possibility will be interesting to test with future observations.

\section{Summary}
We have performed pulse-phase spectroscopy of SMC X-1 from several
epochs and have found that: 
\begin{enumerate}
\im The spectra are well fit by a soft blackbody plus hard power-law
spectrum, and the soft and hard components vary separately with pulse phase.
\im The shapes of pulses in the hard component remain essentially the
same at different epochs, but the soft component pulses change on
timescales $>$ a few days.
\im These varying soft profiles can be explained, in principle, by a
model in which the soft emission comes from reprocessing of the rotating
pulsar beam by a warped inner accretion disk that precesses with time.
Precession of such a disk is believed to cause the known superorbital periodicity in SMC X-1.
\im Further observations are required to establish whether the long-term changes in the hard and
soft pulses of SMC X-1 are in fact correlated to the superorbital
period, and to determine the precise disk and beam geometries.
\end{enumerate}

\acknowledgements 
We are grateful to Joseph Neilsen for his contribution to the data analysis,
Ramesh Narayan for valuable discussions, and to the referee for
constructive comments. This work was supported by NSF grant AST
0307433, and has made use of NASA's Astrophysics Data System, and of
data obtained from the High Energy Astrophysics Science Archive
Research Center (HEASARC), provided by NASA's Goddard Space Flight
Center.

%\pagebreak
%REFERENCES

%\newpage
%FIGURES

%\bibliographystyle{apj}
%\bibliography{research}

\begin{thebibliography}{55}
\expandafter\ifx\csname natexlab\endcsname\relax\def\natexlab#1{#1}\fi

\bibitem[{{Becker}(1998)}]{beck98}
{Becker}, P.~A. 1998, \apj, 498, 790

\bibitem[{{Becker} \& {Wolff}(2005)}]{beck05}
{Becker}, P.~A. \& {Wolff}, M.~T. 2005, ApJ in press, astro-ph/0505129

\bibitem[{{Blum} \& {Kraus}(2000)}]{blum00}
{Blum}, S. \& {Kraus}, U. 2000, \apj, 529, 968

\bibitem[{{Burderi} {et~al.}(2000){Burderi}, {Di Salvo}, {Robba}, {La Barbera},
  \& {Guainazzi}}]{burd00}
{Burderi}, L., {Di Salvo}, T., {Robba}, N.~R., {La Barbera}, A., \&
  {Guainazzi}, M. 2000, \apj, 530, 429

\bibitem[{{Clark} {et~al.}(1994){Clark}, {Woo}, \& {Nagase}}]{clar94}
{Clark}, G.~W., {Woo}, J.~W., \& {Nagase}, F. 1994, \apj, 422, 336

\bibitem[{{Clarkson} {et~al.}(2003){Clarkson}, {Charles}, {Coe}, {Laycock},
  {Tout}, \& {Wilson}}]{clar03}
{Clarkson}, W.~I., {Charles}, P.~A., {Coe}, M.~J., {Laycock}, S., {Tout},
  M.~D., \& {Wilson}, C.~A. 2003, \mnras, 339, 447

\bibitem[{{Davis}(2001)}]{davi01}
{Davis}, J.~E. 2001, ApJ, 562, 575

\bibitem[{{Deeter} {et~al.}(1998){Deeter}, {Scott}, {Boynton}, {Miyamoto},
  {Kitamoto}, {Takahama}, \& {Nagase}}]{deet98}
{Deeter}, J.~E., {Scott}, D.~M., {Boynton}, P.~E., {Miyamoto}, S., {Kitamoto},
  S., {Takahama}, S., \& {Nagase}, F. 1998, \apj, 502, 802

\bibitem[{{Elsner} \& {Lamb}(1977)}]{elsn77}
{Elsner}, R.~F. \& {Lamb}, F.~K. 1977, \apj, 215, 897

\bibitem[{{Endo} {et~al.}(2000){Endo}, {Nagase}, \& {Mihara}}]{endo00}
{Endo}, T., {Nagase}, F., \& {Mihara}, T. 2000, \pasj, 52, 223

\bibitem[{{Ghosh} \& {Lamb}(1978)}]{ghos78}
{Ghosh}, P. \& {Lamb}, F.~K. 1978, \apjl, 223, L83

\bibitem[{{Ghosh} {et~al.}(1977){Ghosh}, {Pethick}, \& {Lamb}}]{ghos77}
{Ghosh}, P., {Pethick}, C.~J., \& {Lamb}, F.~K. 1977, \apj, 217, 578

\bibitem[{{Gruber} \& {Rothschild}(1984)}]{grub84}
{Gruber}, D.~E. \& {Rothschild}, R.~E. 1984, \apj, 283, 546

\bibitem[{{Hickox} {et~al.}(2004){Hickox}, {Narayan}, \& {Kallman}}]{hick04}
{Hickox}, R.~C., {Narayan}, R., \& {Kallman}, T.~R. 2004, \apj, 614, 881

\bibitem[{{Hutchings} {et~al.}(1977){Hutchings}, {Cowley}, {Osmer}, \&
  {Crampton}}]{hutc77}
{Hutchings}, J.~B., {Cowley}, A.~P., {Osmer}, P.~S., \& {Crampton}, D. 1977,
  \apj, 217, 186

\bibitem[{{Jones} \& {Forman}(1976)}]{jone76}
{Jones}, C. \& {Forman}, W. 1976, \apjl, 209, L131

\bibitem[{{Katz}(1973)}]{katz73}
{Katz}, J.~I. 1973, \nat, 246, 87

\bibitem[{{Kohno} {et~al.}(2000){Kohno}, {Yokogawa}, \& {Koyama}}]{kohn00}
{Kohno}, M., {Yokogawa}, J., \& {Koyama}, K. 2000, \pasj, 52, 299

\bibitem[{{Kraus} {et~al.}(1996){Kraus}, {Blum}, {Schulte}, {Ruder}, \&
  {Meszaros}}]{krau96}
{Kraus}, U., {Blum}, S., {Schulte}, J., {Ruder}, H., \& {Meszaros}, P. 1996,
  \apj, 467, 794

\bibitem[{{Kretschmar} {et~al.}(1997){Kretschmar}, {Pan}, {Kendziorra},
  {Maisack}, {Staubert}, {Skinner}, {Pietsch}, {Truemper}, {Efremov}, \&
  {Sunyaev}}]{kret97}
{Kretschmar}, P., {Pan}, H.~C., {Kendziorra}, E., {Maisack}, M., {Staubert},
  R., {Skinner}, G.~K., {Pietsch}, W., {Truemper}, J., {Efremov}, V., \&
  {Sunyaev}, R. 1997, \aap, 325, 623

\bibitem[{{Lang} {et~al.}(1981){Lang}, {Levine}, {Bautz}, {Hauskins}, {Howe},
  {Primini}, {Lewin}, {Baity}, {Knight}, {Rotschild}, \& {Petterson}}]{lang81}
{Lang}, F.~L., {Levine}, A.~M., {Bautz}, M., {Hauskins}, S., {Howe}, S.,
  {Primini}, F.~A., {Lewin}, W.~H.~G., {Baity}, W.~A., {Knight}, F.~K.,
  {Rotschild}, R.~E., \& {Petterson}, J.~A. 1981, \apjl, 246, L21

\bibitem[{{Leahy}(2002)}]{leah02}
{Leahy}, D.~A. 2002, MNRAS, 334, 847

\bibitem[{{Leahy}(2003)}]{leah03}
---. 2003, \apj, 596, 1131

\bibitem[{{Leong} {et~al.}(1971){Leong}, {Kellogg}, {Gursky}, {Tananbaum}, \&
  {Giacconi}}]{leon71}
{Leong}, C., {Kellogg}, E., {Gursky}, H., {Tananbaum}, H., \& {Giacconi}, R.
  1971, \apjl, 170, L67+

\bibitem[{{Levine} {et~al.}(1996){Levine}, {Bradt}, {Cui}, {Jernigan},
  {Morgan}, {Remillard}, {Shirey}, \& {Smith}}]{levi96}
{Levine}, A.~M., {Bradt}, H., {Cui}, W., {Jernigan}, J.~G., {Morgan}, E.~H.,
  {Remillard}, R., {Shirey}, R.~E., \& {Smith}, D.~A. 1996, \apjl, 469, L33+

\bibitem[{{Liller}(1973)}]{lill73}
{Liller}, W. 1973, \apjl, 184, L37+

\bibitem[{{Lucke} {et~al.}(1976){Lucke}, {Yentis}, {Friedman}, {Fritz}, \&
  {Shulman}}]{luck76}
{Lucke}, R., {Yentis}, D., {Friedman}, H., {Fritz}, G., \& {Shulman}, S. 1976,
  \apjl, 206, L25

\bibitem[{{Lutovinov} {et~al.}(2004){Lutovinov}, {Tsygankov}, {Grebenev},
  {Pavlinsky}, \& {Sunyaev}}]{luto04}
{Lutovinov}, A.~A., {Tsygankov}, S.~S., {Grebenev}, S.~A., {Pavlinsky}, M.~N.,
  \& {Sunyaev}, R.~A. 2004, Astronomy Letters, 30, 50

\bibitem[{{Meszaros} \& {Riffert}(1988)}]{mesz88}
{Meszaros}, P. \& {Riffert}, H. 1988, \apj, 327, 712

\bibitem[{{Nagase}(1989)}]{naga89}
{Nagase}, F. 1989, \pasj, 41, 1

\bibitem[{{Naik} \& {Paul}(2002)}]{naik02}
{Naik}, S. \& {Paul}, B. 2002, Journal of Astrophysics and Astronomy, 23, 27

\bibitem[{{Naik} \& {Paul}(2003)}]{naik03}
---. 2003, \aap, 401, 265

\bibitem[{{Naik} \& {Paul}(2004{\natexlab{a}})}]{naik04a}
---. 2004{\natexlab{a}}, \aap, 418, 655

\bibitem[{{Naik} \& {Paul}(2004{\natexlab{b}})}]{naik04}
---. 2004{\natexlab{b}}, \apj, 600, 351

\bibitem[{{Neilsen} {et~al.}(2004){Neilsen}, {Hickox}, \& {Vrtilek}}]{neil04}
{Neilsen}, J., {Hickox}, R.~C., \& {Vrtilek}, S.~D. 2004, \apjl, 616, L135

\bibitem[{{Paul} {et~al.}(2002){Paul}, {Nagase}, {Endo}, {Dotani}, {Yokogawa},
  \& {Nishiuchi}}]{paul02}
{Paul}, B., {Nagase}, F., {Endo}, T., {Dotani}, T., {Yokogawa}, J., \&
  {Nishiuchi}, M. 2002, \apj, 579, 411

\bibitem[{{Pravdo} {et~al.}(1979){Pravdo}, {White}, {Szymkowiak}, {Boldt},
  {Holt}, {Serlemitsos}, {Swank}, {Tuohy}, \& {Garmire}}]{prav79}
{Pravdo}, S.~H., {White}, N.~E., {Szymkowiak}, A.~E., {Boldt}, E.~A., {Holt},
  S.~S., {Serlemitsos}, P.~J., {Swank}, J.~H., {Tuohy}, I., \& {Garmire}, G.
  1979, \apj, 231, 912

\bibitem[{{Preciado} {et~al.}(2002){Preciado}, {Boroson}, \&
  {Vrtilek}}]{prec02}
{Preciado}, M.~E., {Boroson}, B., \& {Vrtilek}, S.~D. 2002, \pasp, 114, 340

\bibitem[{{Ramsay} {et~al.}(2002){Ramsay}, {Zane}, {Jimenez-Garate}, {den
  Herder}, \& {Hailey}}]{rams02}
{Ramsay}, G., {Zane}, S., {Jimenez-Garate}, M.~A., {den Herder}, J., \&
  {Hailey}, C.~J. 2002, \mnras, 337, 1185

\bibitem[{{Reynolds} {et~al.}(1993){Reynolds}, {Parmar}, \& {White}}]{reyn93}
{Reynolds}, A.~P., {Parmar}, A.~N., \& {White}, N.~E. 1993, \apj, 414, 302

\bibitem[{{Romanova} {et~al.}(2003){Romanova}, {Ustyugova}, {Koldoba}, {Wick},
  \& {Lovelace}}]{roma03}
{Romanova}, M.~M., {Ustyugova}, G.~V., {Koldoba}, A.~V., {Wick}, J.~V., \&
  {Lovelace}, R.~V.~E. 2003, \apj, 595, 1009

\bibitem[{{Schreier} {et~al.}(1972){Schreier}, {Giacconi}, {Gursky}, {Kellogg},
  \& {Tananbaum}}]{schr72}
{Schreier}, E., {Giacconi}, R., {Gursky}, H., {Kellogg}, E., \& {Tananbaum}, H.
  1972, \apjl, 178, L71+

\bibitem[{{Scott} {et~al.}(2000){Scott}, {Leahy}, \& {Wilson}}]{scot00}
{Scott}, D.~M., {Leahy}, D.~A., \& {Wilson}, R.~B. 2000, \apj, 539, 392

\bibitem[{{Tananbaum} {et~al.}(1972){Tananbaum}, {Gursky}, {Kellogg},
  {Levinson}, {Schreier}, \& {Giacconi}}]{tana72}
{Tananbaum}, H., {Gursky}, H., {Kellogg}, E.~M., {Levinson}, R., {Schreier},
  E., \& {Giacconi}, R. 1972, \apjl, 174, L143+

\bibitem[{{Unger} {et~al.}(1992){Unger}, {Norton}, {Coe}, \& {Lehto}}]{unge92}
{Unger}, S.~J., {Norton}, A.~J., {Coe}, M.~J., \& {Lehto}, H.~J. 1992, \mnras,
  256, 725

\bibitem[{{Vrtilek} {et~al.}(1994){Vrtilek}, {Mihara}, {Primini}, {Kahabka},
  {Marshall}, {Agerer}, {Charles}, {Cheng}, {Dennerl}, {La Dous}, {Hu},
  {Rutten}, {Serlemitsos}, {Soong}, {Stull}, {Truemper}, {Voges}, {Wagner}, \&
  {Wilson}}]{vrti94}
{Vrtilek}, S.~D., {Mihara}, T., {Primini}, F.~A., {Kahabka}, P., {Marshall},
  H., {Agerer}, F., {Charles}, P.~A., {Cheng}, F.~H., {Dennerl}, K., {La Dous},
  C., {Hu}, E.~M., {Rutten}, R., {Serlemitsos}, P., {Soong}, Y., {Stull}, J.,
  {Truemper}, J., {Voges}, W., {Wagner}, R.~M., \& {Wilson}, R. 1994, \apjl,
  436, L9

\bibitem[{{Vrtilek} {et~al.}(2001){Vrtilek}, {Raymond}, {Boroson}, {Kallman},
  {Quaintrell}, \& {McCray}}]{vrti01}
{Vrtilek}, S.~D., {Raymond}, J.~C., {Boroson}, B., {Kallman}, T., {Quaintrell},
  H., \& {McCray}, R. 2001, \apjl, 563, L139

\bibitem[{{Vrtilek} {et~al.}(2005){Vrtilek}, {Raymond}, {Boroson}, \&
  {McCray}}]{vrti05}
{Vrtilek}, S.~D., {Raymond}, J.~C., {Boroson}, B., \& {McCray}, R. 2005, \apj,
  626, 307

\bibitem[{{Webster} {et~al.}(1972){Webster}, {Martin}, {Feast}, \&
  {Andrews}}]{webs72}
{Webster}, B.~L., {Martin}, W.~L., {Feast}, M.~W., \& {Andrews}, P.~J. 1972,
  Nature Physical Science, 240, 183

\bibitem[{{Wijers} \& {Pringle}(1999)}]{wije99}
{Wijers}, R.~A.~M.~J. \& {Pringle}, J.~E. 1999, \mnras, 308, 207

\bibitem[{{Wojdowski} {et~al.}(1998){Wojdowski}, {Clark}, {Levine}, {Woo}, \&
  {Zhang}}]{wojd98}
{Wojdowski}, P., {Clark}, G.~W., {Levine}, A.~M., {Woo}, J.~W., \& {Zhang},
  S.~N. 1998, \apj, 502, 253

\bibitem[{{Woo} {et~al.}(1995){Woo}, {Clark}, {Blondin}, {Kallman}, \&
  {Nagase}}]{woo95}
{Woo}, J.~W., {Clark}, G.~W., {Blondin}, J.~M., {Kallman}, T.~R., \& {Nagase},
  F. 1995, \apj, 445, 896

\bibitem[{{Woo} {et~al.}(1996){Woo}, {Clark}, {Levine}, {Corbet}, \&
  {Nagase}}]{woo96}
{Woo}, J.~W., {Clark}, G.~W., {Levine}, A.~M., {Corbet}, R.~H.~D., \& {Nagase},
  F. 1996, \apj, 467, 811

\bibitem[{{Yokogawa} {et~al.}(2000){Yokogawa}, {Paul}, {Ozaki}, {Nagase},
  {Chakrabarty}, \& {Takeshima}}]{yoko00}
{Yokogawa}, J., {Paul}, B., {Ozaki}, M., {Nagase}, F., {Chakrabarty}, D., \&
  {Takeshima}, T. 2000, \apj, 539, 191

\bibitem[{{Zane} {et~al.}(2004){Zane}, {Ramsay}, {Jimenez-Garate}, {Willem den
  Herder}, \& {Hailey}}]{zane04}
{Zane}, S., {Ramsay}, G., {Jimenez-Garate}, M.~A., {Willem den Herder}, J., \&
  {Hailey}, C.~J. 2004, \mnras, 350, 506

\end{thebibliography}

%\newpage
%TABLES
%Missions and Observations of SMC X-1

%Disk and beam parameters

% 12.0       & 16.9       & 16.7       & 4.0       &      1.1  \\
% 10.6       & 15.4       & 13.8       & 3.5       &      1.8  \\
% 10.1       & 15.2       & 13.8       & 3.1       &      1.0  \\
% 11.8       & 16.5       & 15.4       & 3.6       &      1.8  \\
% 12.6       & 17.6       & 16.9       & 3.9       &      2.4  \\
% \W6.1      & \W7.3      & \W6.3      & \bf{0.6}  &      2.4  \\     
% \bf{\W4.8} & \bf{\W6.2} & \bf{\W5.3} & 1.4       &      2.6  \\
% 12.8       & 17.5       & 17.5       & 4.0       &      0.7  

\end{document}